\documentclass[twocolumn,showpacs,prb,floatfix,superscriptaddress,
citeautoscript ,cite]{revtex4}
\usepackage{graphicx}
\usepackage{color}
\usepackage{amsmath}

\usepackage{booktabs}

\begin{document}

\title{Pentagonal Monolayer Crystals of Carbon, Boron Nitride, and Silver Azide}

\author{M. Yagmurcukardes} 
\email{mehmetyagmurcukardes@iyte.edu.tr}
\affiliation{Department of Physics, Izmir Institute of Technology, 35430 Izmir, 
Turkey}

\author{H. Sahin}
\affiliation{Department of Physics, University of Antwerp, 2610, Antwerp, 
Belgium}

\author{J. Kang}
\affiliation{Department of Physics, University of Antwerp, 2610, Antwerp, 
Belgium}

\author{E. Torun}
\affiliation{Department of Physics, University of Antwerp, 2610, Antwerp, 
Belgium}

\author{F. M. Peeters}
\affiliation{Department of Physics, University of Antwerp, 2610, Antwerp, 
Belgium}

\author{R. T. Senger}
\email{tugrulsenger@iyte.edu.tr }
\affiliation{Department of Physics, Izmir Institute of Technology, 35430 Izmir, 
Turkey}

\date{\today}


\begin{abstract}

In this study we present a theoretical investigation of structural, electronic 
and mechanical properties of pentagonal monolayers of carbon (p-graphene), 
boron nitride (p-B$_{2}$N$_{4}$ and p-B$_{4}$N$_{2}$) and silver azide 
(p-AgN$_{3}$) by performing state-of-the-art first principles calculations. 
Our total energy calculations suggest feasible formation of monolayer crystal 
structures composed 
entirely of pentagons. In addition, 
electronic band dispersion calculations indicate that while p-graphene and 
p-AgN$_{3}$ are  semiconductors with indirect bandgaps, p-BN structures 
display metallic  behavior. We also investigate the mechanical properties 
(in-plane stiffness and the Poisson's ratio) of four different pentagonal structures 
under uniaxial strain. p-graphene is found to have the highest stiffness 
value and the corresponding Poisson's ratio is found to be negative. Similarly, 
p-B$_{2}$N$_{4}$ and p-B$_{4}$N$_{2}$ have negative Poisson's ratio  
values. On the other hand,  the p-AgN$_{3}$  has a large and positive 
Poisson's ratio. In dynamical stability tests based on calculated phonon 
spectra of these pentagonal monolayers, we find that only p-graphene and 
p-B$_{2}$N$_{4}$ are stable, but p-AgN$_{3}$ and p-B$_{4}$N$_{2}$ are 
vulnerable against vibrational excitations. 

\end{abstract}

\maketitle

\section{Introduction}

In the last decade, graphene, one atom thick form of carbon atoms arranged in a 
honeycomb  structure, has become one of the most exciting topics of materials 
research due to its exceptional properties\cite{Novo1,Geim1}. Besides graphene 
\cite{Novo2},  there exists many other forms of pure carbon in nature such as 
graphite, diamond, C$_{60}$ fullerene\cite{Kroto}, nanotube\cite{Iijima}, carbon 
nanocone\cite{Charlier}, nanochain\cite{Jin} and graphdiyne\cite{Li} which are 
the well known bulk and low dimensional forms of carbon element. In 
additon to these, stability and unique mechanical properties of a new carbon 
allotrope, p-graphene, are reported by Zhang et al. recently\cite{S.Zhang}. 
It is shown that while the unique pentagonal crystal symmetry provides a 
dynamical stability (for temperatures up to 1000 K), the buckled nature of the 
p-graphene leads to a negative value for its Poisson's ratio.

The synthesis of graphene\cite{Novo2,Berger} made other two  dimensional 
materials, such as hexagonal structures of III-V binary 
compounds\cite{Novo3,hasan1}, a popular field of research. Moreover, 
one-dimensional forms of AlN and BN, as nanotubes and nanoribbons were studied 
before\cite{Zhang,H.Park,Topsakal,Barone,Zhao,Zhukovskii}.  Hexagonal monolayer 
structures of these compounds, for example, $h$-BN  \cite{Zeng,Song} and 
$h$-AlN\cite{Bacaksiz,Novo3,Zhuang,QWang,KKim,MFarahani} are wide band-gap 
semiconductors with a nonmagnetic ground state. Recently the synthesis of 
$h$-AlN by Tsipas et al.\cite{Tsipas} motivated further study of the properties 
of $h$-AlN. Very recently we have reported unique thickness-dependent features 
of the electronic structure of $h$-AlN crystal.\cite{Bacaksiz} 

Metal azides, consisting of a metal atom (Na, K, Rb, Cs, Ag, Cu or Tl) and the 
azide molecule (N$_{3}$), are another group of compounds which may find 
applications in monolayer crystal technology. Their electronic structure, 
chemical bonding, vibrational and optical properties have been 
investigated\cite{Gordienko1,Jain,Aluker,Gordienko2,Gordienko3,Aduev,W.Zhu1, 
W.Zhu2, W.Zhu3,Evans1,Evans2,Colton,Schmidt,Hou}. Due to its large chemical 
energy stored in its bulk phases, AgN$_{3}$ is one of the intensely studied 
members of this family. Gordienko et al.\cite{Gordienko1} have studied the 
electronic band structure of AgN$_{3}$ by using density functional theory (DFT) 
calculations. Jain et al.\cite{Jain} calculated the energy band gap of AgN$_{3}$ 
as 2.95 eV. In addition, in the study by Aluker et al.\cite{Aluker} the chemical 
bonding between the Ag and N atoms were studied by using a pseudopotential 
approach. Using a pseudoatomic orbital basis, the electronic structure of 
AgN$_{3}$ was also reported\cite{Gordienko2}. Change of structural 
and vibrational properties of AgN$_{3}$ under applied pressure was studied by 
using DFT and generalized gradient approximation (GGA)\cite{W.Zhu1}. 
Moreover, Schmidt  et al.\cite{Schmidt} reported the crystal structure and 
chemical bonding of the high temperature phase of AgN$_{3}$ by using X-ray 
powder diffraction. In this study it was pointed out that the high 
temperature-AgN$_{3}$ phase contains buckled layers with silver atom connecting 
to the azide groups in pentagonal form in the direction parallel to [001]. The 
phase transitions and structures of AgN$_{3}$ at different pressure values were 
also reported by Hou et al.\cite{Hou} 

In this study we investigate the structural, electronic and mechanical 
properties of pentagonal monolayers of carbon (p-graphene), two phases of 
boron nitride (p-B$_{2}$N$_{4}$ and p-B$_{4}$N$_{2}$) and silver azide 
(p-AgN$_{3}$). The mechanical properties of these pentagonal structures are 
examined under uniaxial strain and in terms of the in-plane stiffness and the 
Poisson's ratio values. Their vibrational spectra are also calculated. The 
paper is organized as follows: The details of our 
computational methodology are given in Sec. \ref{comp}. Structural properties of 
four different pentagonal structures are presented in Sec. \ref{structural}. The 
electronic and magnetic properties of optimized structures are investigated in 
Sec. \ref{selectro}. In Sec. \ref{mechanical} mechanical properties and 
dynamical stability of the pentagonal structures are investigated. Finally we 
present our conclusions in Sec.\ref{conc}.
 
\begin{figure}
\includegraphics[width=8.5cm]{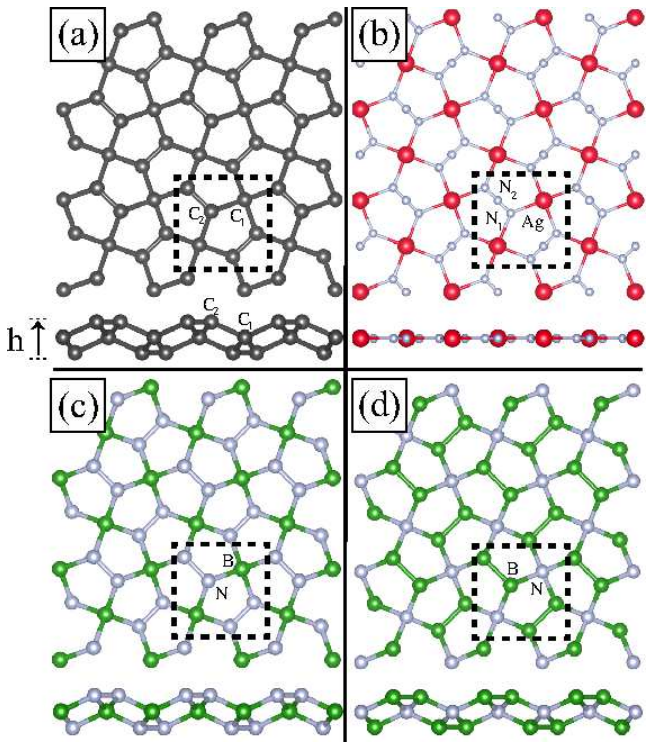}
\caption{\label{stuc}
(Color online) Top view and side view of pentagonal (a) graphene (b) AgN$_{3}$, (c) 
B$_{2}$N$_{4}$, (d)  B$_{4}$N$_{2}$.}
\end{figure}

\section{Computational Methodology}\label{comp}

In this study the first-principles calculations were performed within the 
framework of density functional theory (DFT) by using the Vienna Ab initio 
Simulation Package (VASP) package\cite{vasp1,vasp2,vasp3,vasp4}. The  approach 
is based on an iterative solution of the Kohn-Sham equations\cite{Kohn-Sham} 
with a plane-wave set adopted with the Perdew-Burke-Ernzerhof (PBE) 
exchange-correlation functional of the generalized gradient approximation (GGA) 
\cite{GGA-PBE1,GGA-PBE2}. In order to analyze the  charge transfers the Bader 
technique was used\cite{Henkelman}.

Electronic and geometric relaxations of the pentagonal structures of the 
monolayers  were performed by considering the following criteria in our 
calculations. The energy cut-off value for plane wave basis set was taken to be 
$500$ eV. The global break condition for the electronic self consistent-loop was 
considered to be $10^{-5}$ eV. For geometric relaxation of the structures a  
unit cells containing 6 or 8 atoms were. For this purpose the minimum energy was 
calculated by varying the lattice constant values and the pressure in all 
directions is decreased to a value smaller than 1 kbar. Brillouin zone 
integration was performed by using a set of $5\times5\times1$ Gamma-centered 
k-point sampling mesh. For density of states and work function calculations a 
set of $15\times15\times1$ k-point sampling was used to get more accurate 
results. The cohesive energy of a unit cell was calculated using the formula 
$E_c$=$\sum$$n_a$ $E_a$ $-$ $E_{str}$, where $E_a$ denotes the energy 
of a single isolated atom and $n_a$ denotes the number of atoms 
contained in the unit cell. $E_{str}$ denotes the 
total energy of the monolayer structure. Summation is used for the structure containing 
different types of atoms in its simulation cell. Calculated cohesive energies
are listed in Table \ref{table}.

\begin{table*}
\caption{\label{table} Geometry of pentagonal structures, calculated lattice 
parameter $a$, the distance between atoms $d$$_{XY}$, buckling of the monolayer $h$, 
total magnetic moment $\mu$, the 
amount of charge lost or gained by the atoms $\Delta\rho$, the total cohesive 
energy of a primitive unitcell $E_c$, 
the energy band gap of the structure 
$E$$_g$, work function $\Phi$, Poisson's ratio $\nu$ and in-plane stiffness 
$C$.}
\begin{tabular}{rcccccccccccccccc}
\hline\hline
& Geometry & $a$ & $d_{XY}$  & $h$ & $\mu$ & $\Delta\rho$ 
& $E_c$ & $E_g$ & $\Phi$ & $\nu$ & $C$ &\\
& & (\AA{})& (\AA{}) & (\AA{}) & ($\mu{_B}$) & ($e$) & (eV) & (eV) & (eV) & & 
(eV/\AA$^2$)&\\
\hline
p-Graphene & $buckled$ & 3.64 & 1.34 (C$_{1}$-C$_{1}$) & 1.21 & 0 & 0.3 & 42.40 
& 2.21 & 6.01 & $-0.08$ & 16.71 \\
&    &      & 1.55 (C$_{1}$-C$_{2}$) &      &      &      &  \\
p-AgN$_{3}$ & $planar$ & 6.01 & 1.19 (N-N) &-&0& 2.1 & 31.45 & 1.33 & 3.43 
& 0.90 & 0.37 \\
&  &  & 2.33 (Ag-N) &   &  &  &   \\
p-B$_{2}$N$_{4}$ & $buckled$ & 3.62 & 1.34 (N-N)& 1.26 & 0 & 4.2 & 34.49 & - & 
5.19 & $-0.02$ & 3.62 \\
&  &  & 1.55 (B-N) &   &  &  &   \\
p-B$_{4}$N$_{2}$ & $buckled$ & 3.79 & 1.59 (B-B)  & 1.23 & 1.95  & 4.3 & 33.58 & - & 3.88 & $-0.19$ & 7.59 &\\
 &  &  & 1.57 (N-B) &  &  &  &   \\
\hline\hline
\end{tabular}
\end{table*}

\section{Structural Properties}\label{structural}

Firstly geometrical relaxations of structures were performed by considering 
their square-shaped primitive unitcells 
with the lattice vectors $a_{1}$=$a$(1,0,0) and $a_{2}$=$a$(0,1,0) 
for 
all 
structures (see Fig. \ref{stuc}).  
In the structure of p-graphene the 4-coordinated carbon atoms were denoted 
by 
C$_{1}$ while 
the 3-coordinated ones were denoted by C$_{2}$. The geometrical calculations 
show that the bond length of 
C$_{1}$-C$_{2}$ is 1.55 \AA{} while C$_{2}$-C$_{2}$ bond length is 1.34 \AA{}. 
The lattice constant is $a$=3.64 \AA{} 
within GGA approximation and it is consistent with the value calculated by 
Zhang et al.\cite{S.Zhang} 
The buckling of the layer is 1.21 \AA{} which is also consistent with the value 
calculated by Zhang et al.\cite{S.Zhang} 
Bader charge analysis indicates 
that 0.3 $e$ amount of charge is donated from  C$_{1}$ and two C$_{2}$ 
atoms to other two C$_{2}$ atoms. The calculated cohesive energy is 42.40 
eV for 
p-graphene monolayer.

For p-AgN$_{3}$ geometry relaxation 8-atomic primitive unitcell was 
considered. As seen in Fig. \ref{stuc}(b)  
2-coordinated N atoms are denoted by N$_{1}$ while 3-coordinated ones are 
denoted by N$_{2}$. The 
geometry relaxation within the GGA approximation gives the lattice constant 
value 
as $a$=6.02 \AA{}. The Ag-N$_{1}$ bond length is 
2.33 \AA{} while the N$_{1}$-N$_{2}$ bond length is 1.19 \AA{}. The bond angle 
between the Ag-N$_{1}$-Ag atoms is 132.5 
degrees and it is 90 degrees for the N$_{1}$-Ag-N$_{1}$ bonds. The relaxed 
geometry 
of AgN$_{3}$ monolayer structure is  
planar similar to some other two dimensional structures such as hexagonal 
graphene and h-BN. Bader charge analysis 
shows that an amount of 0.7 $e$ charge from each Ag atom was donated to the 
N atoms but dominantly to the central ones. The final charge on the Ag, 
N$_{1}$ and N$_{2}$ atoms and 
N$_{1}$ atom are 10.3 $e$, 5.2 $e$ and 5.3 $e$ 
respectively. 
The total cohesive energy of p-AgN$_{3}$ is 31.45 eV as listed in Table \ref{table}.

Optimized lattice constant of the p-B$_{2}$N$_{4}$ is found to be $a$=3.62 
\AA{}. The N-N and B-N bond lengths are 1.34 \AA{} and 1.55 \AA{}, 
respectively. The buckling of p-B$_{2}$N$_{4}$ is 1.26 \AA{} which is close 
to that of p-graphene. The Bader charge analysis demonstrates that B atoms have 
final charge of 0.9 $e$ so that an amount of 2.1 $e$ charge was transferred 
to the N atoms from each B atom. The cohesive energy of p-B$_{2}$N$_{4}$ monolayer is calculated as 34.49 eV.

The p-B$_{4}$N$_{2}$ has a lattice constant of  $a$=3.79 \AA{} which is greater 
than 
that of p-B$_{2}$N$_{4}$. This time the B-N bond length is 1.57 \AA{} while 
the B-B bond length is 1.59 \AA{}. The buckling of p-B$_{4}$N$_{2}$ is 
1.23 \AA{} which is close to that of the p-B$_{2}$N$_{4}$ structure. 
Results of 
Bader charge analysis indicates that an amount of 2.2 $e$ charge was depleted to 
each N atom from the B atoms. Finally the cohesive energy of 
p-B$_{4}$N$_{2}$ 
is 
33.58 eV.

\section{Electronic Properties}\label{selectro}

\begin{figure}
\includegraphics[width=8.5cm]{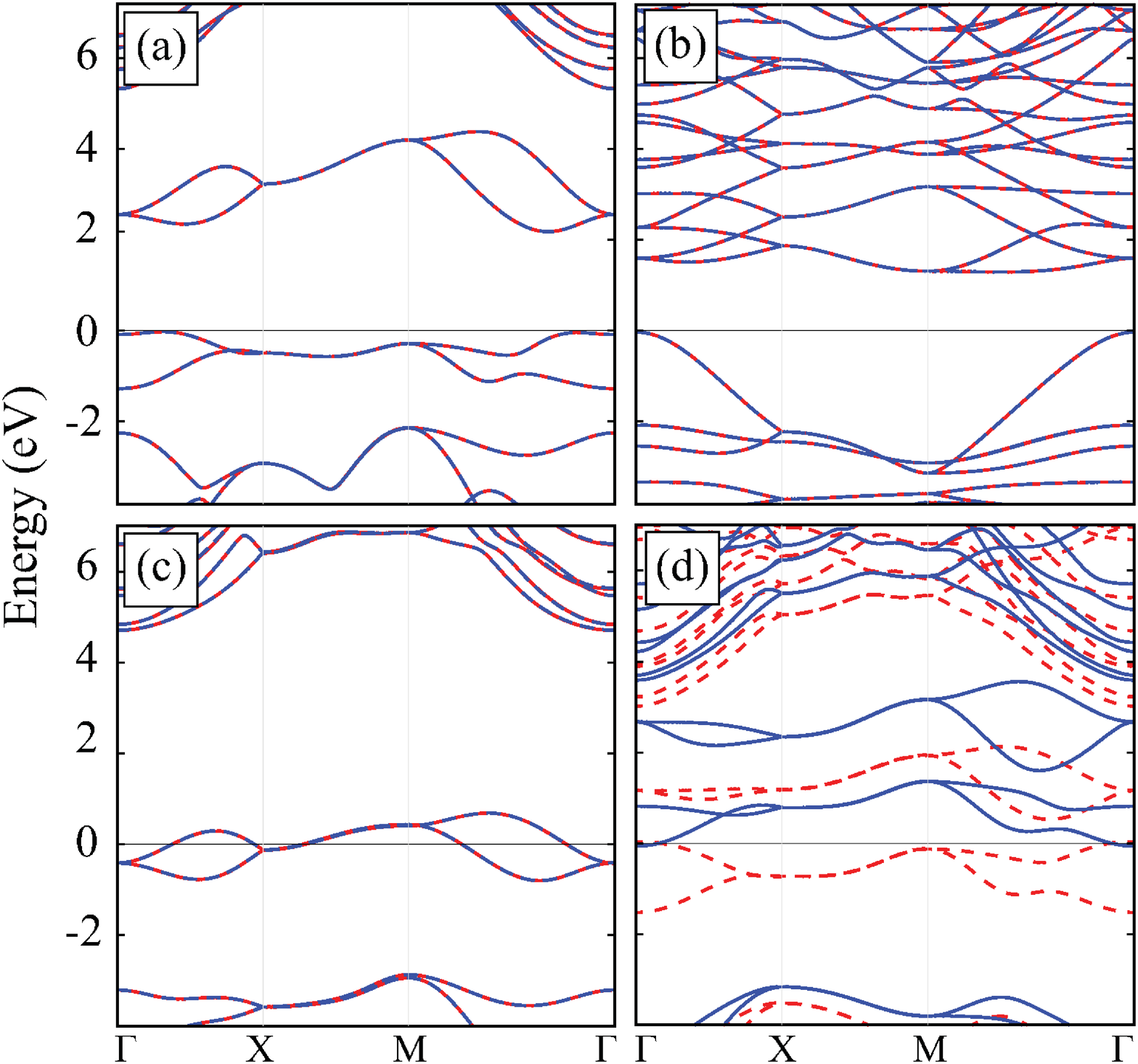}
\caption{\label{bands}
(Color online) Band-structures of pentagonal (a) graphene (b) AgN$_{3}$, (c) 
B$_{2}$N$_{4}$, (d)  B$_{4}$N$_{2}$ where blue lines denote up spins while 
dashed red lines denote down spins respectively.}
\end{figure}

In this section the electronic band dispersion and magnetic ground state of 
p-graphene, p-AgN$_{3}$, p-B$_{2}$N$_{4}$ and p-B$_{4}$N$_{2}$ 
are investigated comprehensively. 
As seen in Table \ref{table}, the p-graphene has an indirect band gap 
of 2.21 eV. As shown in 
Fig. \ref{bands}(a) the valence band maximum (VBM) of the p-graphene is 
located 
in between the $\Gamma$ and the $X$ (high symmetry) points while the conduction 
band minima (CBM) is in between the $M$ and the $\Gamma$ points. 
It also appears that the both spin up and spin down states are degenerate 
throughout the Brillouin Zone and thus the 
structure does not exhibit any spin polarization in its ground state. 
In the 6-atomic primitive unitcell of the p-graphene while two of the 
4-coordinated C atoms have no excess electrons, four 3-coordinated C atoms pair 
their electrons in $p_{z}$ orbitals and therefore the p-graphene has a 
nonmagnetic ground state.

The p-AgN$_{3}$ has an indirect band gap of 1.33 eV as seen in Table 
\ref{table}.
In Fig. \ref{bands}(b) the VBM of the p-AgN$_{3}$ is in 
between the $\Gamma$ and the $X$ points while the CBM exists in between 
the $M$ and the $\Gamma$ points. As it is seen in Fig. 
\ref{bands}(b) that the p-AgN$_{3}$ also does not 
exhibit any spin.

The p-B$_{2}$N$_{4}$ is another structure having nonmagnetic ground 
state. As seen in 
Fig. \ref{bands}(c), again the spin up and the spin down states are degenerate. 
Unlike the p-graphene and the p-AgN$_{3}$, 
the p-B$_{2}$N$_{4}$ displays metallic behavior. 
The valence band crosses the Fermi level in between all high symmetry points 
through whole Brillouin Zone.

In all the pentagonal structures considered, only the p-B$_{4}$N$_{2}$ has a 
spin 
polarization in its ground state. The total magnetic moment of 
p-B$_{4}$N$_{2}$ is 1.95 $\mu_B$ as given in Table \ref{table}. This value of total magnetic moment arises 
from the ferromagnetic ordering of B local moments. In the primitive unit cell each B atom 
has a local magnetic moment of 0.48 $\mu_B$ while the each N atom has local moment about 0.02 $\mu_B$  
which is very small compared to that of B atom. Therefore, the net magnetic moment of 1.95 $\mu_B$  for p-B$_{4}$N$_{2}$ 
structure is mostly due to local moments of B atoms. In its 
6-atomic 
primitive unit cell both N atoms are 4-coordinated while all the B atoms are 3 
coordinated. The spin polarization 
is localized on the N atoms since they add up their electrons in their $p_{z}$ 
orbitals. 
As given in Fig. \ref{bands}(d) the spin up and 
spin down states have different dispersions. Only in between the high symmetry 
points $\Gamma$ and the $X$, 
$M$ and the $\Gamma$ the spin up and spin down bands cross each other just 
above the Fermi level. 
The valence band of spin down states crosses  Fermi level while the conduction 
band of spin up states crosses Fermi level. The 
band structure metallic for both spins but if 
spin orbit coupling is included, then there may open a band gap at the points 
where the up and down spin bands cross.

\begin{figure} [htbp]
\includegraphics[width=8.5cm]{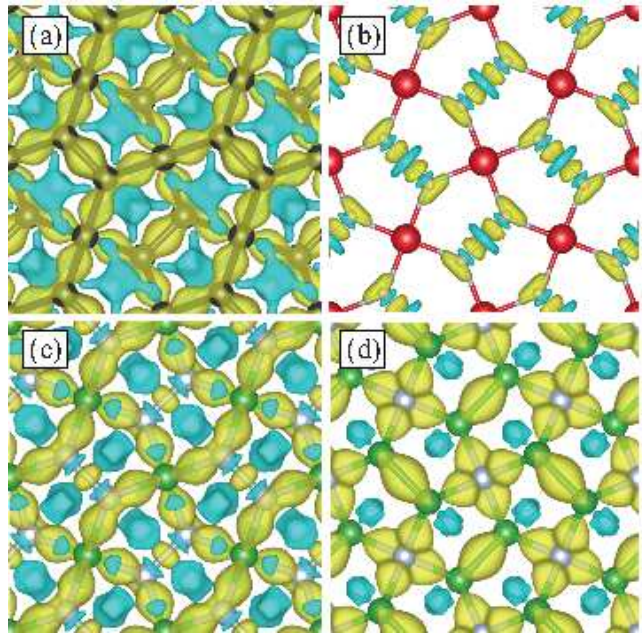}
\caption{\label{chargedendiff}
(Color online){ Charge density difference of pentagonal (a) graphene, (b) AgN$_3$, 
(c) B$_2$N$_4$ and (d) B$_4$N$_2$.} }
\end{figure}

The charge density difference plots of pentagonal structures are provided in 
Fig. \ref{chargedendiff}. In order to plot these figures we first obtained the 
total charge density of each material. Then, using the same unit cell and 
settings we obtained the charge density of each atom separately at their 
original positions in the compound. After that we summed these  
individual charge densities and subtracted them from the charge density of the
compound. These figures reveal the modifications in the total charge of 
the individual atoms when the crystal is formed. The charge density 
difference plot of p-graphene Fig. \ref{chargedendiff}(a) shows 
that there is a charge depletion in the hollow site of the lattice. 
This charge is accumulated mostly at the bonding sited between the C atoms. 
The figure for the AgN$_3$, Fig. \ref{chargedendiff}(b), indicates that there is a 
charge depletion from the N$_2$ atoms and a charge accumulation at the region 
where 
the N$_1$-N$_2$ and N$_1$-Ag chemical bonds are formed. The hollow site charge 
depletion 
is also observed for B$_2$N$_4$ in Fig. \ref{chargedendiff}(c). 
Similar to previous cases, there is a charge accumulation at the locations 
where 
the B-N chemical bonds are formed. For the case of B$_4$N$_2$ in Fig. 
\ref{chargedendiff}(d), there is a charge depletion from the one side of the B 
atoms and again a charge accumulation at the bonding sites. 

For the p-graphene, the charge transfer is from C$_1$ atoms and 2 of  C$_2$ 
atoms to other C$_2$ 
atoms. For the 
p-AgN$_{3}$, as seen in Fig. 
\ref{chargedendiff}(b) there exists a charge depletion from Ag and N$_1$ 
atoms to central N atoms in azide group. For the
p-B$_2$N$_4$ structure all of the charge given in Table 
\ref{table} is 
depleted to the N atoms as depicted by the charge density plot in Fig. 
\ref{chargedendiff}(c). Finally for p-B$_4$N$_2$ 
monolayer again the charge depletion occurs from B atoms to N atoms.

\section{Mechanical Properties}\label{mechanical}
The stiffness can be 
explained as the rigidity or the flexibility of a material. 
The parameter which shows the mechanical response of a material to an applied 
stress
is called the Poisson's ratio. It is defined as the ratio 
of the transverse contraction strain to the longitudinal extension strain in the 
direction of 
stretching force. The in-plane stiffness and the Poisson's ratio can be deduced 
from the relationship between the strain and 
the total energy. To calculate the mentioned parameters, we apply strain 
${\varepsilon}_x$ and 
${\varepsilon}_y$ to these materials by changing the lattice constants along x 
and y directions. The strain range is from -0.02 to 0.02 with a step of 0.01 
which gives a data grid of 25 points.
At each grid point, the atomic positions are relaxed  and the strain energy 
E$_S$, which is the energy difference between 
strained and unstrained structures, is calculated. In the harmonic region the 
strain energy can be fitted as 
$E_S=c_1{{\varepsilon}_x}^2+c_2{{\varepsilon}_y}^2+c_3{\varepsilon}_x{ 
\varepsilon}_y$. The in-plane stiffness along $x$ and $y$ directions can then 
be calculated as $C_x=(1/S_0)(2c_1-{c_3}^2/2c_2)$ and 
$C_y=(1/S_0)(2c_2-{c_3}^2/2c_1)$ where S$_0$ is the unstretched area of the 
supercell. The Poisson's ratio along  $x$ and $y$ 
directions can be obtained by $\nu_x=c_3/2c_2$ and $\nu_y=c_3/2c_1$, 
respectively. For all 
pentagonal structures we find that the in-plane stiffness and the Poisson's ratio 
along x and y directions are equal.

The calculated in-plane stiffness and Poisson's ratio are listed in Table 
\ref{table}. It can be seen that p-graphene has the largest in-plane 
stiffness of 16.71 
eV/{\AA}$^2$, indicating strong bonding between carbon atoms. However, this 
value is smaller than that of graphene, which has an in-plane stiffness of 20.91 
eV/{\AA}$^2$.\cite{Topsakal} This can be attributed to different number of 
bonds in p-graphene and graphene. In graphene, each C atom is
3-fold coordinated, while in graphyne the average coordination
number of C atom is 2.67. P-graphene has fewer number of bonds than 
graphene, so it has relatively smaller in-plane stiffness. The calculated 
Poisson's ratio for p-graphene
is -0.08 which is consistent with the value calculated by Zhang et 
al.\cite{S.Zhang} 

The p-AgN$_{3}$ has a large Poisson's ratio 
of 0.90, revealing its strong ability to preserve the equilibrium area when 
strain is applied. The Poisson's ratio for p-B$_4$N$_2$ is -0.19, 
this is consistent with the calculation of in-plane stiffness. 
P-B$_2$N$_4$ has an in-plane stiffness 
of 3.62 eV/{\AA}$^2$, much smaller than the p-graphene. For 
p-B$_4$N$_2$, the in-plane stiffness is 7.59 eV/{\AA}$^2$. It is 
interesting to note that the p-graphene, the p-B$_2$N$_4$ and 
the p-B$_4$N$_2$ have negative Poisson's ratio values, contrary to the most 
of the existing materials. 
Therefore, they belong to the so-called auxetic structures. When uniaxial 
tensile strain is applied to 
these structures, the lattice along the transverse direction expands rather than 
compresses. 
Normally, this ratio is positive and  most of the solids expand in the 
transverse direction
when they are subjected to a uniaxial compression. The materials with negative 
Poisson's 
ratio unfolds when they are stretched. Therefore, they are isotropic in two 
dimensions for certain lengths and angles.
It has been reported that some artificial materials have negative Poisson's 
ratio and they 
exhibit excellent mechanical properties \cite{Burns,Jiang}. In contrast to 
structure-engineered bulk auxetics,
 the negative Poisson's ratio is intrinsic in single layers of p-graphene, 
p-B$_2$N$_4$ and p-B$_4$N$_2$. 

\begin{figure}
\includegraphics[width=8.5cm]{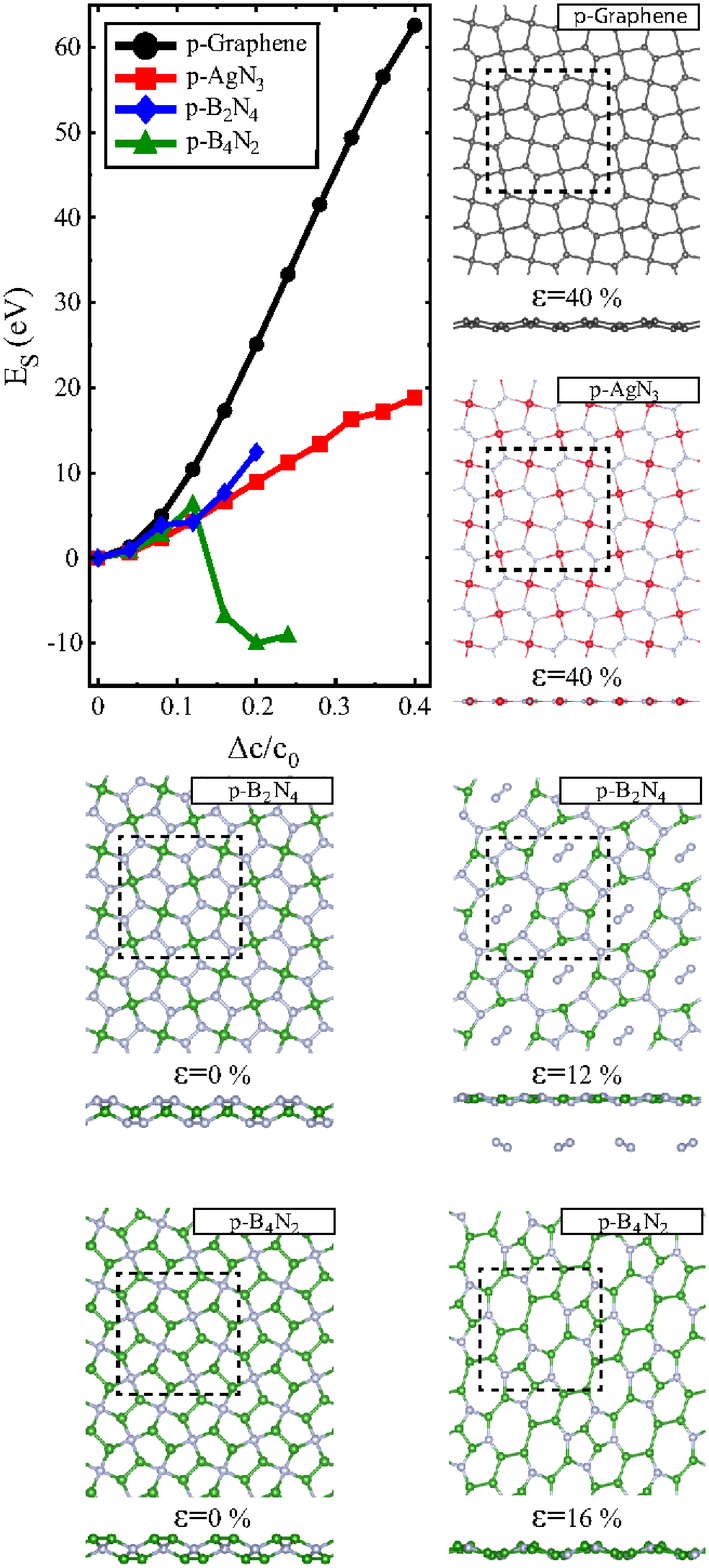}
\caption{\label{mech}
(Color online) Energy variation of the pentagonal structures under applied strain and 
the corresponding atomic configurations at given strain strengths.}
\end{figure}
 
We also consider higher values of strain from 0.04 to 0.40 in uniform
expansion, in order to see structural deformations 
and determine the elastic and plastic regions for each pentagonal structure. For 
this purpose, we prefer a fully symmetric square lattice with well defined high 
symmetry points
in the BZ. Again the calculations are performed in a $2\times2$  
supercell. Increasing the strength of applied strain, increases the total 
energy of the structure. The p-graphene has no structural deformation up to 
the strain 
value of 40\% but the buckling of the layer decreases to 0.66 \AA {}. Under 40\% 
strain, the C$_{2}$-C$_{2}$ and C$_{1}$-C$_{2}$ bond lengths 
are 1.35 \AA {} and 2.15 \AA {}, respectively. P-AgN$_{3}$ also does not 
have any 
structural deformation up to 40\% strain. It remains in the same form 
but with a higher Ag-N$_{1}$ bond length of 3.47 \AA {} while the bond lengths 
in azide group remain the same. The situation is different for pentagonal structures 
of B and N, because they both have deformations in their structures 
at some critical strain values. P-B$_2$N$_4$ has not a pentagonal shape 
structure when 12\% strain is applied. Therefore one may say that it is the 
critical strain value for p-B$_2$N$_4$ between elastic and plastic regions.
Plastic region refers to a region in which irreversible
structural changes occur in the system and it transforms into
a different structure. This critical strain value is slightly greater for 
p-B$_4$N$_2$. After the strain strength of 16\%, p-B$_4$N$_2$ transforms 
into a different structure.

\begin{figure}
\includegraphics[width=8.5cm]{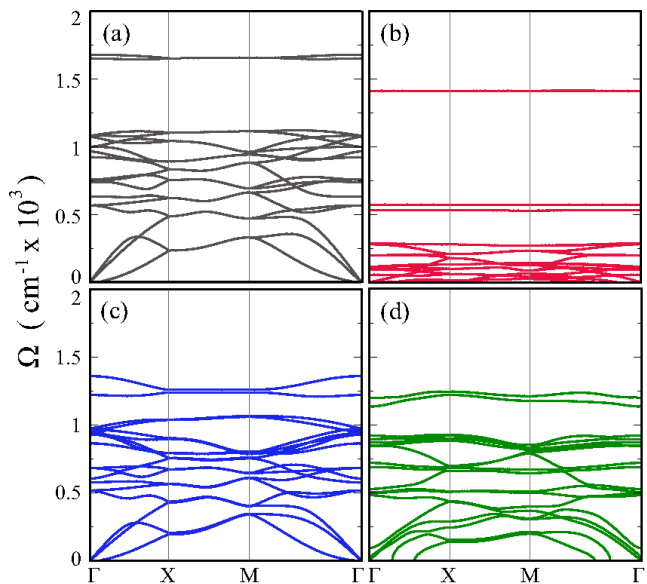}
\caption{\label{phonon}
(Color online) Phonon modes of pentagonal (a) graphene (b) AgN$_{3}$, (c) 
B$_{2}$N$_{4}$, (d)  B$_{4}$N$_{2}$.}
\end{figure}

As an important feature of mechanical properties we also 
examine the dynamical stability 
of pentagonal monolayer structures by performing phonon calculations. 
Here, the dynamical matrix and the vibrational modes were calculated using the small-displacement method 
(SDM)\cite{alfe} with forces obtained from VASP. As shown in Fig. \ref{phonon}, 
while pentagonal structures of graphene and B$_2$N$_4$ have real vibrational 
eigenfrequencies in the whole Brillouin zone,
p-AgN$_{3}$ and p-B$_4$N$_2$ have some phonon branches with zero-frequency 
modes at several points in the Brillouin zone. This is an indication of 
irreversible deformations that can be induced by those vibrational modes. It 
appears that although the total energy calculations yield optimized atomic 
structures of p-AgN$_{3}$ and p-B$_4$N$_2$ these structures are 
dynamically unstable. 
Our calculations also reveal that p-graphene and p-B$_2$N$_4$ not only possess dynamically stable crystal 
structures but also have quite high-frequency phonon modes indicating strong bond formation in these materials.

\section{Conclusions}\label{conc}
Motivated by the unique properties of the recently reported p-graphene we 
have investigated the structural, mechanical and electronic properties of three 
novel pentagonal structures as well as p-graphene. Our calculations demonstrate 
that pentagonal 
structures of graphene and BN have buckled geometries while p-AgN$_{3}$ has 
a planar geometry. Calculated band structures show that although hexagonal 
graphene is a zero-band gap semiconductor, the band dispersion of p-graphene 
displays an indirect-band-gap semiconductor behavior. Also the band dispersion 
of 
p-AgN$_{3}$ displays semiconducting behavior with an indirect band gap. 
However, pentagonal structures of BN are metallic while hexagonal 
BN monolayer is a 
wide-band-gap semiconductor. For all of the pentagonal structures investigated 
in this 
study only p-B$_4$N$_2$ has a magnetic ground state while the other 
structures have nonmagnetic ground states. We have also studied the mechanical 
properties of these structures and calculated their in-plane stiffness and 
corresponding Poisson's ratios. The stiffest monolayer is found to be the 
p-graphene among the four structures. p-graphene, p-B$_2$N$_4$ and p-B$_4$N$_2$ 
all have 
negative Poisson's ratio while the p-AgN$_{3}$ has a positive Poisson's 
ratio. Also the uniform strain calculations indicate that p-graphene and 
p-AgN$_{3}$ do not show any irreversible structural deformations for up to 
large strain 
values while p-B$_2$N$_4$ and p-B$_4$N$_2$ deform into different phases at some 
certain strain strengths.

\begin{acknowledgments}

This work was supported by the Flemish Science Foundation (FWO-Vl) and the 
Methusalem foundation of the Flemish government. Computational resources were 
provided by TUBITAK ULAKBIM, High Performance and Grid Computing Center (TR-Grid 
e-Infrastructure). H.S. is supported by a FWO Pegasus Long Marie Curie 
Fellowship. H.S. and R.T.S. acknowledge the support from 
TUBITAK through project 114F397.

\end{acknowledgments}


\begin{thebibliography}{99}


-----------------------------------------

\bibitem{Novo1} K. S. Novoselov, A. K. Geim, S. V. Morozov, D. Jiang, Y. Zhang, 
S. V.
Dubonos, I. V. Grigorieva, and A. A. Firsov,
Science \textbf{306}, 666 (2004).

\bibitem{Geim1}K. S. Novoselov, A. K. Geim, S. V. Morozov, D. Jiang, M. I. 
Katsnelson,
I. V. Grigorieva, S. V. Dubonos, and A. A. Firsov,
Nature \textbf{438}, 197 (2005).

\bibitem{Novo2} 
    K. S. Novoselov, A. K. Geim, S. V. Morozov, D. Jiang, Y. Zhang, S. V. 
Dubonos, I. V. Grigorieva, and A. A. Firsov, 
    Science \textbf{306}, 666 (2004).

\bibitem{Kroto} H. W. Kroto, J. R. Heath, S. C. O'Brien, R. F. Curl, R. E. Smalley, 
Nature \textbf{318}, 162 (1985).

\bibitem{Iijima} S. Iijima, T. Ichihashi, Nature \textbf{363}, 603 (1993).

    
\bibitem{Charlier} J. C. Charlier, G. M. Rignanese, Phys. Rev.
Lett. \textbf{86} 5970 (2001). 


\bibitem{Jin} C. Jin, H. Lan, L. Peng, K. Suenaga, S. Iijima, Phys. Rev. 
Lett. \textbf{102}, 205501 (2009).


\bibitem{Li} Y. Li, L. Xu, H. Liu, Y. Li, Chem. Soc. Rev. \textbf{43}, 
2572 (2014). 

\bibitem{S.Zhang} S. Zhanga, J. Zhouc, Q. Wanga, X. Chend, 
Y. Kawazoef, P. Jenac, PNAS \textbf{112}, 2372 
(2015).

\bibitem{Berger} C. Berger, Z. Song, X. Li, X. Wu, N. 
Brown, C. Naud, D. Mayou, T. Li, J. Hass, A. N. Marchenkov,
E. H. Conrad, P. N. First, W. A. de Heer, Science \textbf{312}, 
1191 (2006).


\bibitem{Novo3} K. S. Novoselov, D. Jiang, F. Schedin, T. Booth, V. V. Khot-
kevich, S. Morozov, and A. K. Geim, Proc. Natl. Acad. Science
U.S.A. \textbf{102}, 10451 (2005).

\bibitem{hasan1} H. Sahin, S. Changirov, M. Topsakal, E. Bekaroglu, E. Akturk,
R. T. Senger, and S. Ciraci, Phys. Rev. B \textbf{80}, 155453 (2009).

\bibitem{Zhang} Z. Zhang and W. Guo, Phys. Rev. B \textbf{77} 075403
(2008). 


\bibitem{H.Park}C. H. Park and S. G. Louie, Nano Lett. \textbf{8}, 2200 
(2008).


\bibitem{Barone}  V. Barone and J. E. Peralta, Nano Lett. \textbf{8}, 2210 
(2008).

\bibitem{Topsakal} M. Topsakal, E. Akturk, and S. Ciraci, Phys. Rev. B
\textbf{79}, 115442 (2009).

\bibitem{Zhao} M. W. Zhao, Y. Y. Xia, D. J. Zhang, L. M. Mei, Phys. Rev. B  
\textbf{68} 235415 (2003). 

\bibitem{Zhukovskii} C. Y. F. Zhukovskii, A. I. Popov, C. Balasubramanian, S. 
Bellucci, J. Phys. Condens. Matter  \textbf{18}, S2045 (2006).

\bibitem{Zeng} H. Zeng, H. Zhi, C. Zhang, Z. Wei, X. Wang, X. Guo, W. Bando, Y. 
Golberg, D. Nano Lett. \textbf{10}, 5049
(2010).

\bibitem{Song} L. Song, L. Ci, L. Lu, H. Sorokin, P. B. Jin, C. Ni, J. Kvashnin, 
A. G. Kvashnin, D. G. Lou, J. Yakobson, B. I. Ajayan, P. M. Nano Lett. \textbf{10}, 3209
(2010).

\bibitem{Bacaksiz} C. Bacaksiz, H. Sahin, H. D. Ozaydin, S. Horzum, R. T. 
Senger, and F. M. Peeters, Phys. Rev. B \textbf{91} 085430
(2015). 

\bibitem{Zhuang} H. L. Zhuang and R. G. Hennig, Appl. Phys. Lett. \textbf{101}, 
153109 
(2012).

\bibitem{QWang} Q. Wang, Q. Sun, P. Jena, and Y. Kawazoe, ACS Nano \textbf{3}, 
621
(2009).

\bibitem{KKim} K. K. Kim, A. Hsu, X. Jia, S. M. Kim, Y. Shi, M. Hofmann, D. 
Nezich, J. F. Rodriguez-Nieva,
M. Dresselhaus, T. Palacios, and J. Kong, Nano Lett. \textbf{12}, 161
(2012).

\bibitem{MFarahani} M. Farahani, T. S. Ahmadi, and A. Seif, J. Mol. 
Struct. \textbf{913}, 126
(2009).

\bibitem{Tsipas} P. Tsipas, S. Kassavetis, D. Tsoutsou, E. Xenogiannopoulou, E. 
Golias, S. A. Giamini, C.
Grazianetti, D. Chiappe, A. Molle, M. Fanciulli, and A. Dimoulas, Appl. Phys. 
Lett. \textbf{103}, 251605
(2013).



\bibitem{Gordienko1} A. B. Gordienko, Y. N. Zhuravlev, A. S. Poplavnoi, Phys. 
Stat. Solidi (b) \textbf{198}, 707 (1996). 

\bibitem{Jain} P. Jain, J. Sahariya, H. S. Mund, M. Sharma, B. L. 
Ahuja, Computational Materials Science \textbf{72}, 101 (2013).

\bibitem{Aluker} E. D. Aluker, Y. N. Zhuravlev, V. Y. Zakharov, N. G. Kravchenko, 
V. I. Krasheninin, A. S. Poplavnoi, Russ. Phys. J. \textbf{46}, 855 (2003).

\bibitem{Gordienko2} A. B. Gordienko, A. S. Poplavnoi, Russ. Phys. J. 
\textbf{47}, 1056 (2004). 

\bibitem{W.Zhu1} W. Zhu, H. Xiao, J. Solid State Chem. \textbf{180} 
3521 (2007).

\bibitem{Schmidt} C. L. Schmidt, R. Dinnebier, U. Wedig, M. Jansen, Inorg. Chem.
\textbf{46}, 907 (2007).

\bibitem{Hou} D. Hou, F. Zhang, C. Ji, T. Hannon, H. Zhu, J. Wu, V. I. 
Levitas, Y. Ma, J. App. Phys. \textbf{110}, 023524 (2010).

\bibitem{Gordienko3} A. B. Gordienko, A. S. Poplavnoi, J. Struct. Chem. 
\textbf{46}, 96 (2005).


\bibitem{W.Zhu2} W. Zhu, H. Xiao, J. Compd. Chem. \textbf{29},
176 (2007).

\bibitem{W.Zhu3} W. Zhu, H. Xiao, Struct. Chem. \textbf{21}, 657 (2010).

\bibitem{Evans1} B. L. Evans, A. D. Yoffe, P. Gray, Chem. Rev. \textbf{59} 515
(1959). 

\bibitem{Evans2} B. L. Evans, A. D. Yoffe, Proc. Roy. Soc. London Ser. A 
\textbf{250}, 346 (1959).


\bibitem{Colton}  R. J. Colton, J. W. Rabalaist, J. Chem. Phys. \textbf{64}, 
3481 (1976).


\bibitem{Aduev} B. P. Aduev, E. D. Aluker, G. M. Belokurov, Y. A. Zakharov, A. G. 
Krechetov, J. Exp. Theo. Phys. \textbf{89} 906
(1999). 

\bibitem{vasp1} G. Kresse and J. Hafner, Phys. Rev. B \textbf{47}, 558
(1993).
\bibitem{vasp2} G. Kresse and J. Hafner, Phys. Rev. B \textbf{49}, 14251
(1994).
\bibitem{vasp3} G. Kresse and J. Furthm\"{u}ller, Comput. Mat. Sci. \textbf{6},
15 (1996).
\bibitem{vasp4} G. Kresse and J. Furthm\"{u}ller, Phys. Rev. B \textbf{54}, 11169
(1996).

\bibitem{Kohn-Sham} W. Kohn and L. J. Sham, Phys. Rev. \textbf{140}, A1133 
(1965).

\bibitem{GGA-PBE1} J. P. Perdew, K. Burke, and M. Ernzerhof, Phys. Rev. Lett.
\textbf{77}, 3865 (1996).
\bibitem{GGA-PBE2} J. P. Perdew, K. Burke, and M. Ernzerhof, Phys. Rev. Lett.
\textbf{78}, 1396 (1997).

\bibitem{Henkelman} G. Henkelman, A. Arnaldsson, and H. Jonsson,
Comput Mater Sci \textbf{36}, 354 (2006).

\bibitem{Burns} Burns, Stephen, Science
\textbf{238}, 551 (1987).

\bibitem{Jiang} J. W. Jiang, H. S. Park,
Nat. Commun. \textbf{5}, 4727 (2014).

\bibitem{alfe} D. Alfe , Comput. Phys. Commun. \textbf{180}, 2622 (2009).

\end{thebibliography}
\end{document}